\documentclass[10pt]{article}

\usepackage[OE]{express}
\usepackage{bm}

\newcommand{\pen}{\mbox{pen}}

\newcommand{\argmin}{\rm arg\,min}

\newcommand{\x}{{\bf x}}
\newcommand{\cb}{{\bf c}}

\newcommand{\ellb}{{\boldsymbol{\ell}}}
\newcommand{\y}{{\bf y}}

\newcommand{\A}{{\bf A}}

\usepackage[normalem]{ulem}

\begin{document}
\title{Revealing hidden scenes by photon-efficient occlusion-based opportunistic active imaging}

\author{Feihu Xu,\authormark{1,*,\dag} Gal Shulkind,\authormark{1,2,*} Christos Thrampoulidis,\authormark{1,*} Jeffrey H. Shapiro,\authormark{1,2} Antonio Torralba,\authormark{2,3} Franco N. C. Wong,\authormark{1} and Gregory W. Wornell\authormark{1,2}}

\address{\authormark{1}Research Laboratory of Electronics, Massachusetts Institute of Technology, Cambridge, MA 02139, USA\\
\authormark{2}Department of Electrical Engineering and Computer Science, Massachusetts Institute of Technology, Cambridge, MA 02139, USA\\
\authormark{3}Computer Science and Artificial Intelligence Laboratory, Massachusetts Institute of Technology, Cambridge, MA 02139, USA \\
\authormark{*}These authors contributed equally.\\
\authormark{\dag}Present address: Shanghai Branch, Hefei National Laboratory for Physical Sciences at the Microscale, University of Science and Technology of China, Shanghai 201315, China}
\email{\authormark{\dag}feihuxu@ustc.edu.cn} 


\begin{abstract}
The ability to see around corners, i.e.,  recover details of a hidden scene from its reflections in the surrounding environment, is of considerable interest in a wide range of applications.  However, the diffuse nature of light reflected from typical surfaces leads to mixing of spatial information in the collected light, precluding useful scene reconstruction. Here, we
employ a computational imaging technique that opportunistically exploits the presence of occluding objects, which obstruct probe-light propagation in the hidden scene, to undo the mixing and greatly improve scene recovery. Importantly, our technique obviates the need for the ultrafast time-of-flight measurements employed by most previous approaches to hidden-scene imaging. Moreover, it does so in a photon-efficient manner based on an accurate forward model and a computational algorithm that, together, respect the physics of three-bounce light propagation and single-photon detection. Using our methodology, we demonstrate reconstruction of hidden-surface reflectivity
patterns in a meter-scale environment from non-time-resolved measurements.
Ultimately, our technique represents an instance of a rich and promising new
imaging modality with important potential implications for imaging science.
\end{abstract}

\ocis{ (110.0110) Imaging systems; (110.1758) Computational imaging; (100.3190) Inverse problems.} 


\section{Introduction}
%
In recent years, remarkable advances have 	been achieved in computational imaging, image
processing and computer vision\cite{sun20133d,kirmani2014first,gao2014single,pawlikowska2017single}.
Whereas conventional imaging involves direct line-of-sight transport from a
light source to a scene, and from the scene back to a camera sensor,
the problem of imaging scenes that are hidden from the camera's direct
line of sight, referred to as seeing around corners or non-line-of-sight (NLoS) imaging, has attracted growing interest.
Indeed, the ability to reconstruct hidden scenes has
the potential to be transformative in important and diverse
applications, including, e.g., medicine, transportation,
manufacturing, scientific imaging, public safety, and security.

Techniques for NLoS imaging that have been recently demonstrated
include time-gated viewing from specular reflections \cite{repasi2009advanced,sume2011radar,chakraborty2010multipath,steinvall2011see},
wavefront shaping\cite{mosk2012controlling,katz2012looking}, and
transient imaging, in which time-of-flight (ToF) measurements are
collected \cite{kirmani2009looking,velten2012recovering,gupta2012reconstruction,heide2014diffuse,laurenzis2014nonline,gariepy2015detection,laurenzis2015multiple,buttafava2015non}.  ToF active imaging using short-duration laser pulses (the most commonly used approach) provide only indirect access to scene information, through detection of photons that have been diffusely reflected by intervening surfaces, which mixes the spatial information they carry.  Such systems have used picosecond-resolution ToF
measurements---as obtained from a streak camera \cite{velten2012recovering,gupta2012reconstruction} or a
single-photon avalanche diode (SPAD) detector \cite{gariepy2015detection,laurenzis2015multiple,buttafava2015non}---to recover hidden scenes. However, collecting such measurements involves complicated and costly apparatus \cite{buttafava2015non}. Klein {\em et al.} has reported tracking NLoS objects using intensity images \cite{Klein2016}; however, its tracking problem is parametric in nature, allowing it to retrieve object translation and rotation only in the case of known objects. In contrast, our focus is on a non-parametric setting, with the goal of retrieving the unknown reflectivity pattern on a hidden surface.

Recently, in~\cite{theory2017}, we proposed a new NLoS imaging framework that opportunistically exploits the presence of opaque occluders in the light propagation path within the hidden space to distinguish light emanating from different parts of the hidden scene (Supplementary Movie S1).  This framework was shown to recover spatial information otherwise destroyed by diffuse reflection, \emph{without} reliance on ultrafast ToF measurements. The approach is reminiscent of pin-speck (or anti-pinhole) imaging\cite{cohen1982anti,torralba2012accidental}, in which an occluder in the scene serves as a defacto lens that facilitates imaging. The focus of~\cite{theory2017} was a  theoretical study of the framework. The model developed there assumes additive signal-independent Gaussian noise, hence the reconstruction algorithm and the preliminary experiment reported in~\cite{theory2017} are tailored to a Gaussian-likelihood method. This Gaussian-noise assumption, however, does not adequately represent shot-noise-limited operation, which prevails in the low-photon-count regime.

In this paper, we extend the applicability of occlusion-based NLoS imaging to operation in that low-photon-count regime. We experimentally demonstrate an imaging system with substantially higher photon efficiency than that reported in \cite{theory2017}, performance that is crucial for fast and low-power NLoS imaging. To do so, we develop an accurate forward model and a photon-efficient computational algorithm based on a binomial-likelihood method that, together, respect the physics of three-bounce light propagation and SPAD-based photodetection. As a result, we achieve a 16$\times$ speedup in the data acquisition process, because information from 16$\times$ fewer photon detections than employed in~\cite{theory2017} suffice to produce images of equal quality. Moreover, unlike \cite{theory2017}, we report full details of our experiments that, in addition to the photon-efficiency demonstration, include investigations of issues---such as the effects of occluder size and the algorithm's regularization parameter on scene reconstruction---that were only studied theoretically in \cite{theory2017}.

\begin{figure}[tbh]
  \centering
  \centerline{\includegraphics[width=1\textwidth]{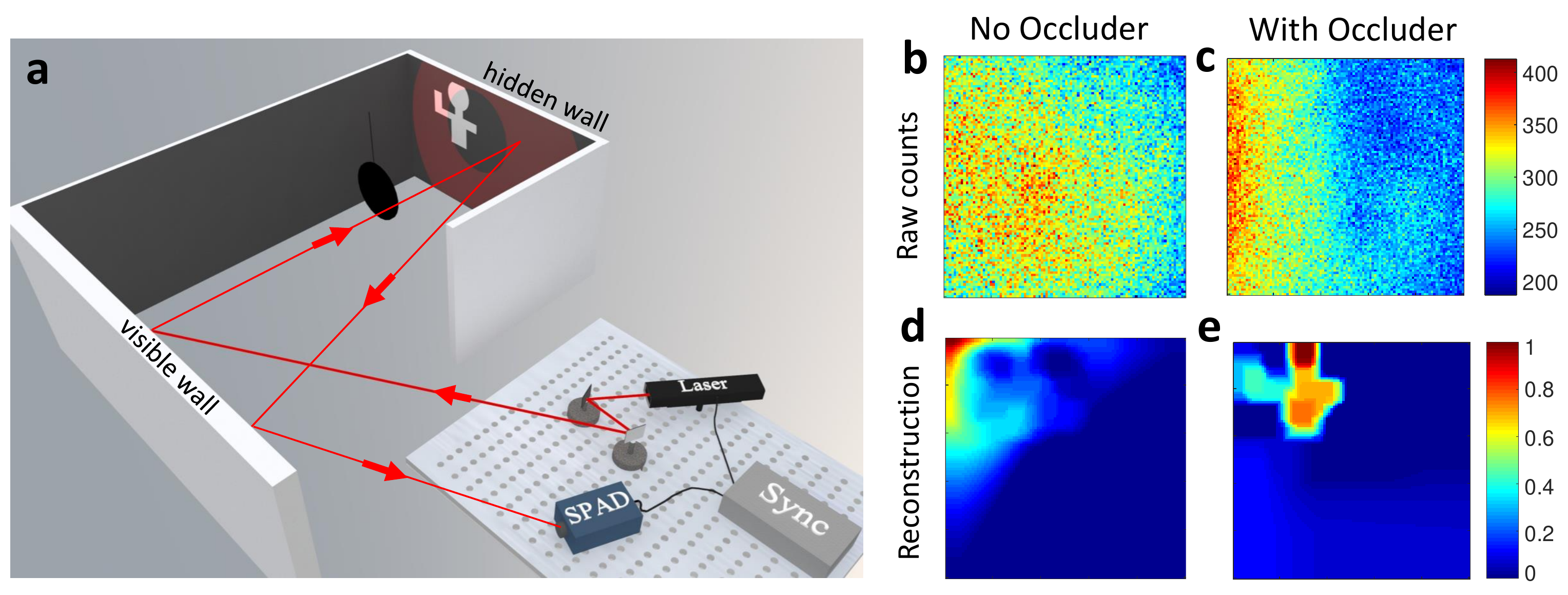}}
  \caption{(a) Experimental configuration. The goal is to reconstruct the reflectivity pattern on the hidden wall. A repetitively-pulsed laser source raster scans a diffuse (nearly Lambertian) visible wall. Photons striking the visible wall reflect toward the hidden wall, reflect at the hidden wall back toward the visible wall, and finally reflect at the visible wall toward the single-photon avalanche diode (SPAD), whose optics are configured to detect backscattered photons from a large patch on the visible wall. The counts are recorded by a single-photon counting module and further computer processed.  When present, an occluder (circular black patch) obstructs some light-propagation paths from the visible wall to the hidden wall (casting a subtle shadow), and from the hidden wall to the visible wall. (b) Raw photon counts in the absence of an occluder. (c) Raw photon counts in the presence of the occluder. (d) Reconstructed reflectivity from the counts in (b). (e) Reconstructed reflectivity from the counts in (c).}
\label{fig:fig1}
\end{figure}

\section{Imaging Scenario}

\begin{figure}[tbh]
  \centering
  \centerline{\includegraphics[width=0.8\textwidth]{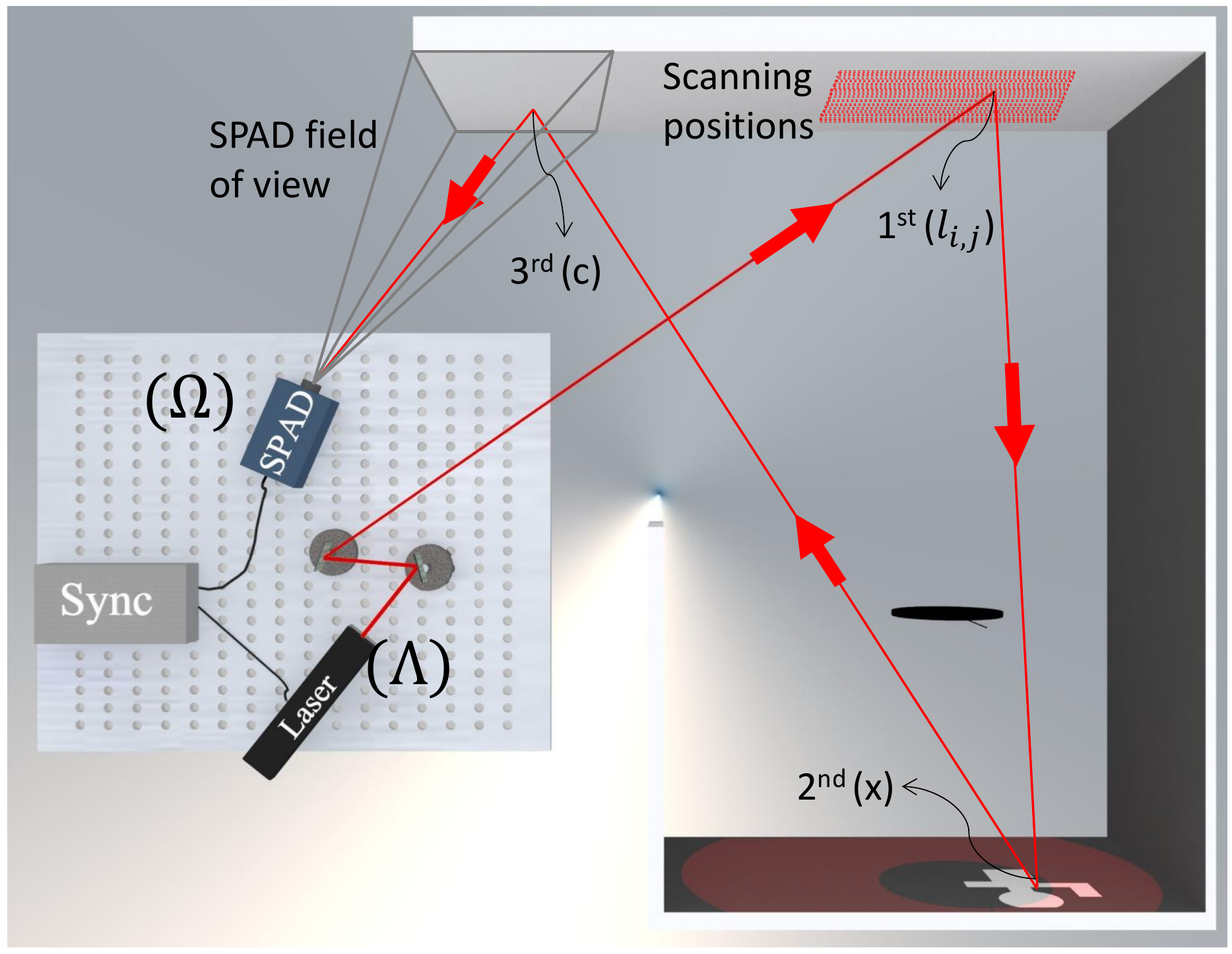}}
  \caption{Top view of experimental setup and a three-bounce light trajectory of the form $\boldsymbol{\Lambda}\rightarrow\ellb_{ij}\rightarrow\x\rightarrow\cb\rightarrow\boldsymbol{\Omega}.$ The laser ($\boldsymbol{\Lambda}$) illuminates the visible wall ($\ellb_{ij}$) and is diffusively reflected (first bounce) toward the hidden wall ($\x$), where it reflects (second bounce) back toward the visible wall. The third-bounce reflection at the visible wall ($\cb$) returns light in the direction of the detector ($\boldsymbol{\Omega}$). A circular occluder is placed between the visible and hidden walls, and partially obstructs light propagating between the visible and hidden walls.} \label{fig:suppfig1}
\end{figure}

Our system configuration is illustrated in Fig.~\ref{fig:fig1}(a) and a top view of the experimental setup is illustrated in Fig.~\ref{fig:suppfig1}. The objective is to reconstruct the unknown reflectivity pattern on the hidden wall. The visible wall
is illuminated by a repetitively-pulsed laser that raster scans an $m\times m$ grid. The photons detected from illumination of a particular scan point have undergone three bounces: first, reflection off the visible wall in the direction of the hidden wall; second, reflection off the hidden wall in the direction of the visible wall, where the reflection is multiplicatively scaled by the reflectivity pattern of the hidden wall we seek to recover; and third, reflection off the visible wall in the direction of a SPAD. As shown in Fig.~\ref{fig:suppfig1}, the SPAD's field of view is configured for the left side of the visible wall, to avoid the direct first bounce and to detect as many third-bounce photons as possible. We use a \emph{single-pixel} SPAD instead of a normal charge-coupled device (CCD) camera, because of its single-photon sensitivity. This is necessary because the returned pulse energy after the three bounces is heavily attenuated ($\sim$130-140 dB in our room-scale experiment), severely limiting the number of detected photons. Thus, the SPAD enables efficient NLoS imaging. We remark that although a SPAD is capable of providing time-stamped measurements, we \emph{discard} the SPAD's time-resolved information by integrating detections over a time-gating window to collect just an $m\times m$ matrix of the raw photon counts obtained from illumination of each laser grid point. To further clarify, we emphasize that a SPAD is not strictly necessary for our imaging method, as we show next that reconstruction is possible when we throw away the detected third-bounce photons' time signatures. Although not demonstrated here, alternative high-sensitivity sensors with no or poor timing resolution---such as an intensified CCD or electron-multiplying CCD---can also be used in our experiment.  We will investigate such modifications in future work.

\begin{figure}[tbh]
  \centering
  \centerline{\includegraphics[width=0.9\textwidth]{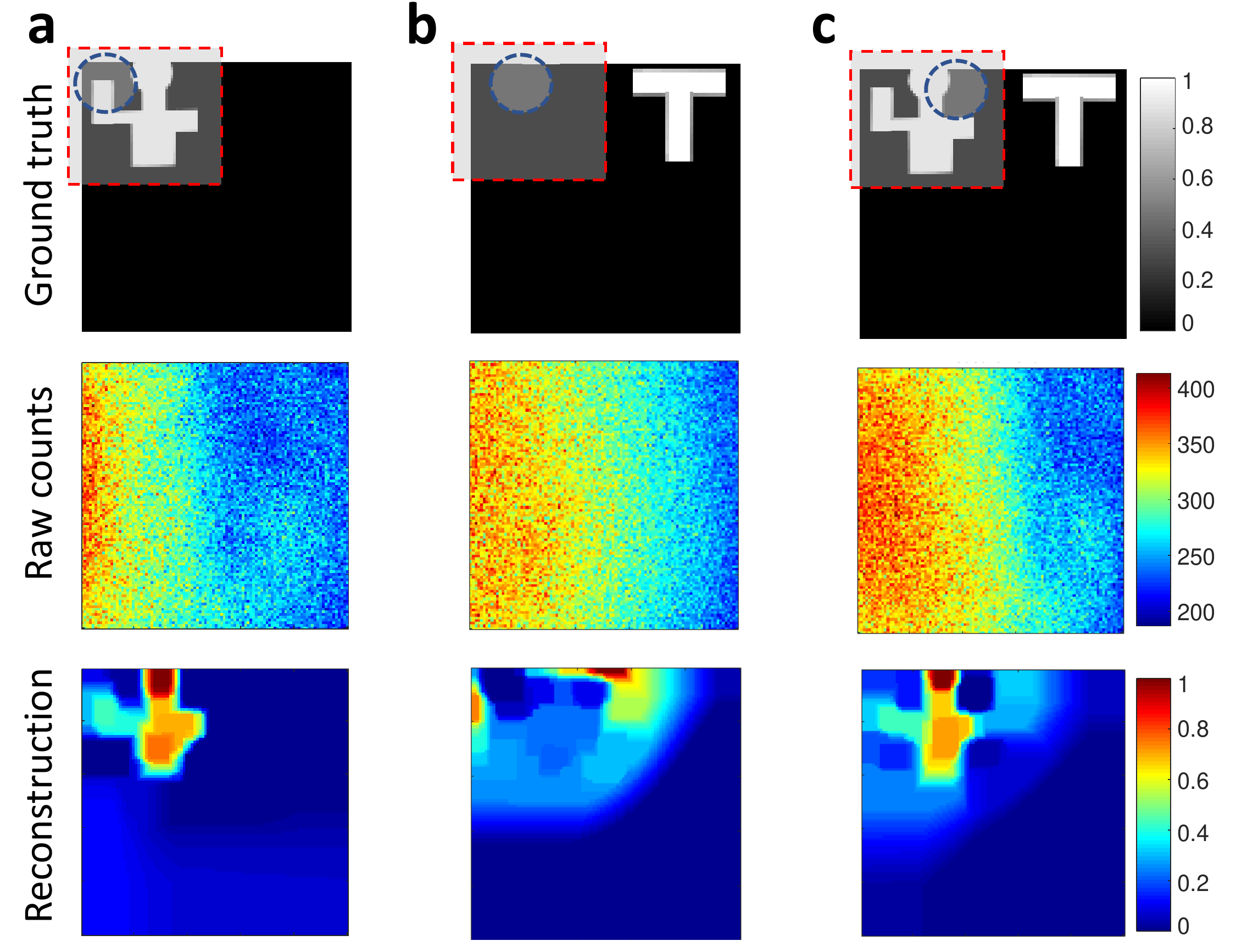}}
\caption{Role of the occluder's shadow in NLoS imaging. The red-dashed square in the ground-truth image indicates the hidden-wall area that is scanned by the occluder's shadow as the laser raster scans the visible wall. The blue-dashed circle in the ground-truth image indicates the approximate occluder-shadow area for one $\ellb_{ij}$ (see Supplementary Movie S2).  (a) The man-shaped pattern, placed in the upper-left quadrant of the hidden wall, is completely scanned by the occluder's shadow pattern as the laser scans the visible wall; with the aid of the occluder, the hidden pattern is successfully reconstructed from the raw counts. (b) The T-shaped pattern, placed in the upper-right quadrant of the hidden wall that is outside of the shadow area, yields raw photon counts that fail to reconstruct the pattern owing to the occluder's shadow not scanning that quadrant. (c) Both the man-shaped pattern and the T-shaped pattern are placed on the hidden wall, with only the man-shaped pattern being scanned by the occluder's shadow, so the man-shaped pattern is reconstructed successfully while the T-shaped pattern is not. } \label{fig:suppfig2}
\end{figure}

We performed this experiment twice, first with no obstruction between the visible and the hidden walls, and then with a black circular occluder inserted between those walls to block some of the light propagating from the visible wall toward the hidden wall, and some of the light propagating from the hidden wall back toward the visible wall, as illustrated in Fig.~\ref{fig:fig1}(a). The corresponding matrices of raw photon counts are shown in Figs.~\ref{fig:fig1}(b) and \ref{fig:fig1}(c). We derive an accurate forward model and solve the resulting inverse problem using a photon-efficient reconstruction algorithm that is tailored to the low-photon-count regime associated with three-bounce propagation (see below). Figures~\ref{fig:fig1}(d) and \ref{fig:fig1}(e) show that reconstruction of the hidden-wall reflectivity pattern failed when photon counts were collected without an occluder being present, but succeeded when they were collected in the presence of the occluder. In Fig.~\ref{fig:suppfig2}, we present experimental results in which different patterns are placed on the hidden wall. These results demonstrate that obstructions in the {light propagation path} enable imaging from non-time-resolved photon counts. Indeed, as will be explained, occluders do so by increasing the informativeness of measurements made in their presence. Note that we have assumed knowledge of the occluder's location.  This information is easily obtained if the occluder is visible from the detector's vantage point.  Moreover, we have initial indication that location information for the occluder can be gleaned from raw-count data (see below).

\section{Forward model}

In this section, we present a ray-optics light propagation model (see details in Appendix~\ref{App:theory}) that relates the unknown reflectivity on a Lambertian hidden wall to the raw photon counts for specified experimental parameters. The model accounts for: (a) third-bounce reflections involving the hidden wall; (b) occlusions in the scene; (c) a low photon-count operating regime; and (d) wide field-of-view detection. For the derivation we assume that:   the geometries of the hidden wall and occluder are known; the occluders are opaque (nonreflecting and nontransmitting); the visible wall is Lambertian with known reflectivity; and the background illumination reaching the detector is known.

The $m\times m$ illumination grid on the visible wall is indexed with
$(i,j)$. The hidden wall is discretized to $n\times n$ pixels indexed
with $(k,l)$. Let ${\bf F}$ be the hidden-wall's \emph{reflectivity matrix}, with entry $0\leq F_{kl}\leq 1$ being the
reflectivity value of the $(k,l)$\/th hidden-wall pixel. We use $Y_{ij}$ to denote the average number of photons arriving at the detector from single-pulse illumination at grid point $(i,j)$, and ${\bf Y}$ to denote the $m\times m$ matrix whose $ij$\/th entry is $Y_{ij}$.  In the absence of background light in three-bounce NLoS
imaging, $Y_{ij}$ is linearly related to ${\bf F}$ as follows (Appendix~\ref{App:theory}):
\begin{equation} \label{eq:linear_measurements_4d}
Y_{ij}= K_{{\rm p}}\sum\limits_{k,l}A^{(ij)}_{kl} F_{kl},
\end{equation}
where $K_{{\rm p}}$ is the average number of photons per transmitted laser
pulse, and $A_{kl}^{(ij)}$, for fixed $i,j$, is the $kl$th entry of an $n\times n$ matrix $\A^{(ij)}$ that is determined by the
physics of light propagation and the geometry of the surfaces
involved. For $1\le i,j\le m$, Eq.~(\ref{eq:linear_measurements_4d}) defines a
linear system of $m^2$ equations in the $n^2$
unknowns ${\bf F}$ that we wish to retrieve.

In practice, ${\bf Y}$ is not directly available. Even were it available, the
robustness of estimating ${\bf F}$ from ${\bf Y}$ would depend on the matrices
$\A^{(ij)}$.  Indeed, high-fidelity inversion of Eq.~(\ref{eq:linear_measurements_4d}) with a finite-precision calculation requires that the $\A^{(ij)}$ vary substantially with $(i,j)$, i.e., that
each laser illumination point retrieves a new informative
projection of the unknowns.
When the space between the visible and hidden walls is free of
obstructions, however, the $\A^{(ij)}$ matrices vary only slightly and smoothly
across different grid points $(i,j)$ (Fig.~\ref{fig:fig1}(b)). Hence
inverting Eq.~(\ref{eq:linear_measurements_4d}) results in
poor reconstruction of the unknown reflectivity because the inversion is ill conditioned
(Fig.~\ref{fig:fig1}(d)).

In contrast, when an occluder is present in the space between the
visible and hidden walls, the matrices $\A^{(ij)}$ in
Eq.~(\ref{eq:linear_measurements_4d}) become much more diverse
(Fig.~\ref{fig:fig1}(c)), enabling much better imaging of the hidden
wall (Fig.~\ref{fig:fig1}(e)). Intuitively, the occluder
partially obstructs light propagation in the hidden space, precluding ${\bf Y}$
contributions from some hidden-scene patches, thus making
some $\A^{(ij)}$ entries vanish.  Moreover,
different laser positions $(i,j)$ and $(i',j')$ may be blocked from illuminating
different portions of the hidden wall (Supplementary Movie S1). Consequently, some
$\A^{(ij)}$ entries that are zeros correspond to $\A^{(i'j')}$ entries that are nonzero, and vice-versa,
yielding the measurement diversity needed for a much better conditioned inversion of Eq.~(\ref{eq:linear_measurements_4d}).

A photon-number-resolving SPAD\cite{hadfield2009single} will produce a Poisson-distributed number of photon counts in response to an illumination pulse\cite{buller2009single}. Currently available SPADs, however, are not
photon-number resolving: after detecting one photon they suffer a dead time\cite{hadfield2009single,buller2009single} whose duration is longer, in our experiment, than the duration of light returned in a single illumination period. Furthermore, after three bounces, the probability of detecting a photon from a single pulse is very low. So, in this low-flux regime, the probability that the SPAD does not detect a photon from single-pulse illumination of
the $ij$th grid point is:
\begin{equation}\label{eq:PoissonP0}
P_0^{(ij)}({\bf F}) = \exp[-\eta( Y_{ij}+B_{ij})],
\end{equation}
where $\eta$ is the SPAD's quantum efficiency, and $B_{ij}$ is the background contribution to the light illuminating the SPAD. Defining $R_{ij}$ to be the number of photons detected from illuminating that grid point with a sequence of $N$ laser pulses, it follows that $R_{ij}$ has a binomial
distribution with success probability $1-P_0^{(ij)}({\bf F})$, i.e.,\cite{shin2015TCI}
\begin{equation}\label{eq:Poisson}
\Pr(R_{ij};{\bf F}) = \left(\begin{array}{c}N \\ R_{ij}\end{array}\right)[1-P_0^{(ij)}({\bf F})]^{R_{ij}}[P_0^{(ij)}({\bf F})]^{N-R_{ij}}.
\end{equation}

\section{Reconstruction algorithm}
To reconstruct the hidden wall's
reflectivity matrix ${\bf F}$ from the $m \times
m$ matrix, ${\bf R}$, of photon counts, we make use of the
forward model from
Eqs.~(\ref{eq:linear_measurements_4d}) to (\ref{eq:Poisson}). In particular, we seek a
matrix $\hat{{\bf F}}$ that maximizes the likelihood
$\mathcal{L}({\bf R};{\bf F})\equiv \prod_{i,j}\Pr(R_{ij};{\bf F})$ of ${\bf F}$ being the true
reflectivity matrix, given that ${\bf R}$ is the observed photon-count matrix. Significantly,
the negative log-likelihood function can be shown to be convex in ${\bf F}$,
and is thus easy to minimize. The optimization program is still convex---and still easily solved---after we impose reflectivity's nonnegativity constraint $F_{kl} \ge 0$, and an additive penalty $\pen({\bf F})$ chosen to ensure spatial correlation between the reflectivity values of neighboring pixels while allowing abrupt reflectivity changes at the boundaries between multipixel regions.
In summary, we reconstruct the reflectivity matrix as the
solution $\hat{{\bf F}}$ to the convex optimization
program
\begin{equation} \label{eq:PML_general}
\hat{{\bf F}}={\argmin}_{{\bf F}\colon F_{kl}\geq 0} \left\{
  -\log[\mathcal{L}({\bf R};{\bf F})] +\lambda\,
  \pen({\bf F}) \right\},
\end{equation}
for an appropriate choice of the regularization parameter $\lambda$. We used the
total-variation (TV) semi-norm penalty function\cite{rudin1992nonlinear}
and a specialized solver\cite{harmany2012spiral} to obtain
$\hat{{\bf F}}$ from Eq.~(\ref{eq:PML_general}).

\begin{figure}[tbh]
  \centering
  \centerline{\includegraphics[width=0.9\textwidth]{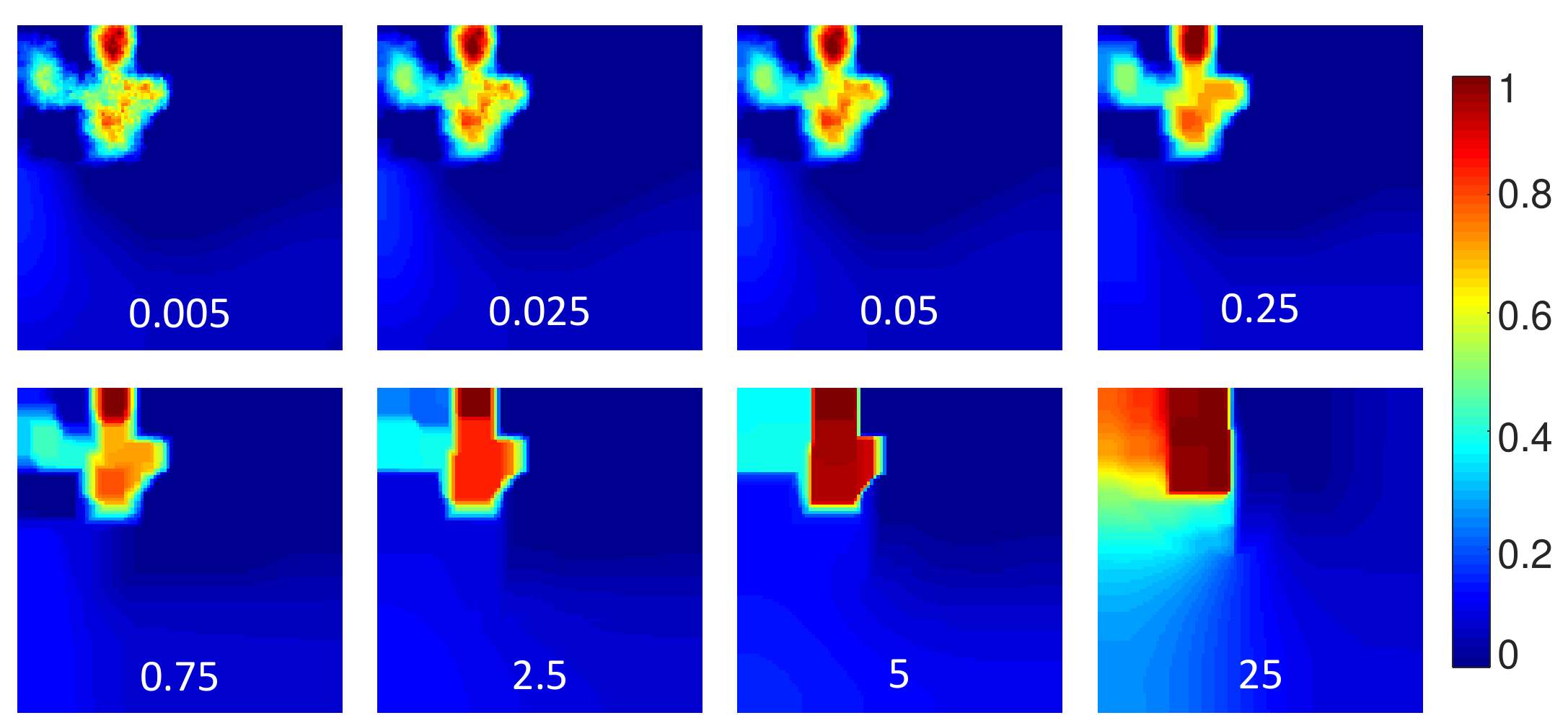}}
\caption{Reconstruction results with different values of the regularization parameter $\lambda$. We demonstrate reconstruction according to Eq.~(\ref{eq:PML_general}) with varying values for the regularization parameter $\lambda$, as indicated on the bottom of the figures. Higher $\lambda$ values promote reconstructions with larger regions of near-uniform reflectivity values, whereas smaller $\lambda$ values produce more detailed but noisier images. In our reconstructions, we chose a $\lambda$ value that does not severely distort the image; here the preferred value is $\lambda=0.75$.} \label{fig:suppfig8}
\end{figure}

The regularization parameter $\lambda$ determines the balance between the two optimization targets:  decreasing the negative log-likelihood and promoting locally-smooth scenes with sharp boundaries. In Fig.~\ref{fig:suppfig8} we demonstrate the effect of varying $\lambda$ on the reconstructed reflectivity. In practice, we choose the regularization parameter to obtain reasonably smooth images that do not seem overly regularized.


\section{Experiment} 

Figure~\ref{fig:suppfig1} depicts the $\sim$1-m scale imaging scene in our experiment. For illumination, we used a repetitively-pulsed 640-nm laser (Picoquant LDH-640B), with sub-ns pulses, 40\,MHz repetition rate, and an average power of $\sim$8\,mW. A two-axis galvo (Thorlabs GVS012) was utilized to raster scan the laser's output over a grid of points (first bounce in Fig.~\ref{fig:suppfig1}) on a nearly-Lambertian visible wall (white poster board, see its characterization in Appendix~\ref{App:experiment}). Light reflected from the visible wall propagates to the hidden wall, where some is reflected back (second bounce) to the visible wall. Finally, some of the second-bounce light that is reflected from the visible wall (third bounce) is collected by a SPAD detector (MPD-PDM with quantum efficiency $\sim$0.35 at 640\,nm). We placed an interference filter (Andover) centered at 640\,nm with a 2\,nm bandwidth in front of the SPAD to suppress background light. The occluder is a nonreflecting black circular patch. In the experiment, the two side walls inside the room were covered with black curtains so that they too are nonreflecting. Note that our forward model can easily take the side walls into consideration were they reflecting. During measurements, we turned off all ambient room light to minimize the background level. In future measurements, one can easily perform the NLoS imaging in the presence of ambient light by operating at an appropriate wavelength outside of the visible range, such as 1550\,nm.

\begin{figure}[tbh]
  \centering
  \centerline{\includegraphics[width=0.9\textwidth]{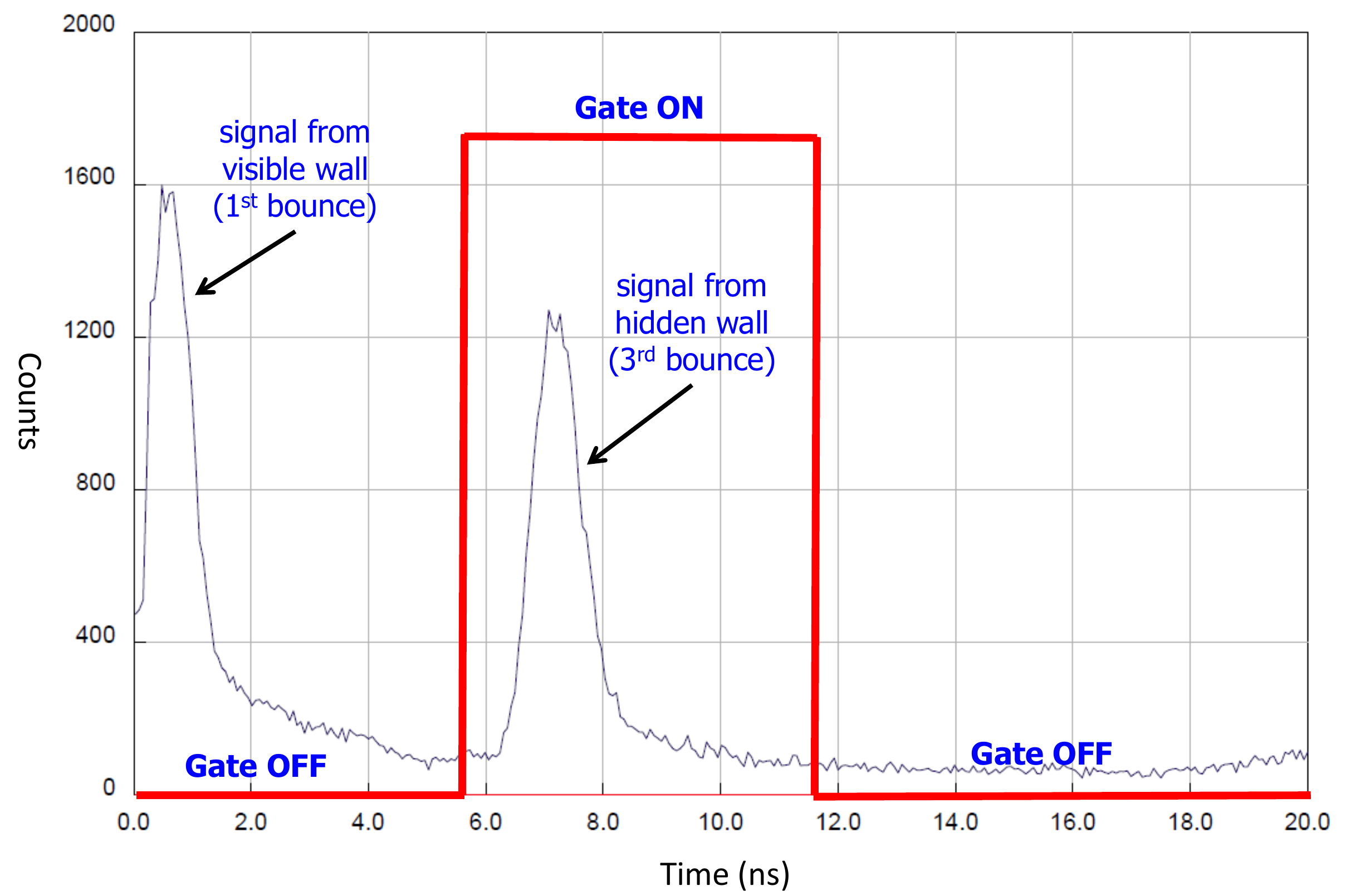}}
\caption{Time-resolved SPAD measurements showing the gated timing window used for post-selecting third-bounce photon detections while suppressing first-bounce photon detections. The gate-off period covers detection times of first-bounce photons and the $\sim$6\,ns duration of the gate-on period is long enough to capture all third-bounce photon detections. In our experiments, only the number of detected photons in gate-on windows were recorded to form the raw-count images.}
\label{fig:suppfig7}
\end{figure}

The focus of the experiment is to utilize the collected third-bounce light to reconstruct the hidden-wall's reflectivity pattern without use of ToF infomation. Therefore it is important to avoid detecting the first-bounce light that will be much stronger than the third-bounce light. We took two initial steps to minimize first-bounce photon detections. First, as shown in Fig.~\ref{fig:suppfig1}, we oriented the SPAD such that its field of view did not overlap the part of the visible wall that was scanned by the laser.  Second, we inserted an opaque screen (not shown in Fig.~\ref{fig:suppfig1}) to block the direct line of sight between the illuminated part of the visible wall and the SPAD.  In testing, however, we found that there was still a substantial number of photon detections from residual first-bounce light, which we could determine from their time delays relative to the laser pulses' emission times. These detections were mostly due to laser light scattered from the two galvo mirrors that illuminated part of the visible wall within the SPAD's field of view. So, because the visible wall is in direct line of sight of the imaging equipment---and hence its location can be easily and accurately estimated---we further suppressed first-bounce photon detections by the following \emph{post-processing} procedure.  We used the time-resolved (TR) information that is automatically captured by the SPAD to set a gated timing window that excludes first-bounce detections but whose duration is long enough to encompass all possible third-bounce detections, as indicated in the measurements shown in Fig.~\ref{fig:suppfig7}. As a result, \emph{no} TR information related to the third-bounce photons is used in our measurements and scene reconstructions. In the future, with better galvo mirrors and a single-photon-sensitive CCD detector, we should be able to perform occlusion-based NLoS imaging with neither the need for nor the possibility of ToF-enabled  suppression of first-bounce photon detections.


\section{Experimental results}
We report experimental results obtained from a meter-scale environment in which the distance between the
detector and the visible wall is $\sim$1.5\,m and a circular occluder of
diameter $\sim$6.8\,cm is positioned roughly midway between the visible
and hidden walls, which are separated by $\sim$1\,m. A $\sim$$0.4\,{\rm m}\times 0.4\,{\rm m}$ reflectivity pattern was mounted on the upper-left quadrant of the $\sim$$1\,{\rm m}\times 1\,{\rm m}$ hidden wall to ensure that the pattern is properly scanned by the occluder's shadow as the laser raster scans the visible wall (see Fig.~\ref{fig:suppfig2} and Supplementary Movie S2). We performed an initial calibration of the background levels $\{B_{ij}\}$ that was then used for all subsequent experiments. We note that the need for background calibration can be avoided with better experimental equipment (see Appendix~\ref{App:experiment}). The occluder's shape and position are assumed to be known for the purpose of scene reconstruction. From the known geometry, the matrix $\A^{(ij)}$ can be determined. Finally, we chose $m=100$ and $n=100$ for our measurements.

\begin{figure}[tbh]
  \centering
  \centerline{\includegraphics[width=0.95\textwidth]{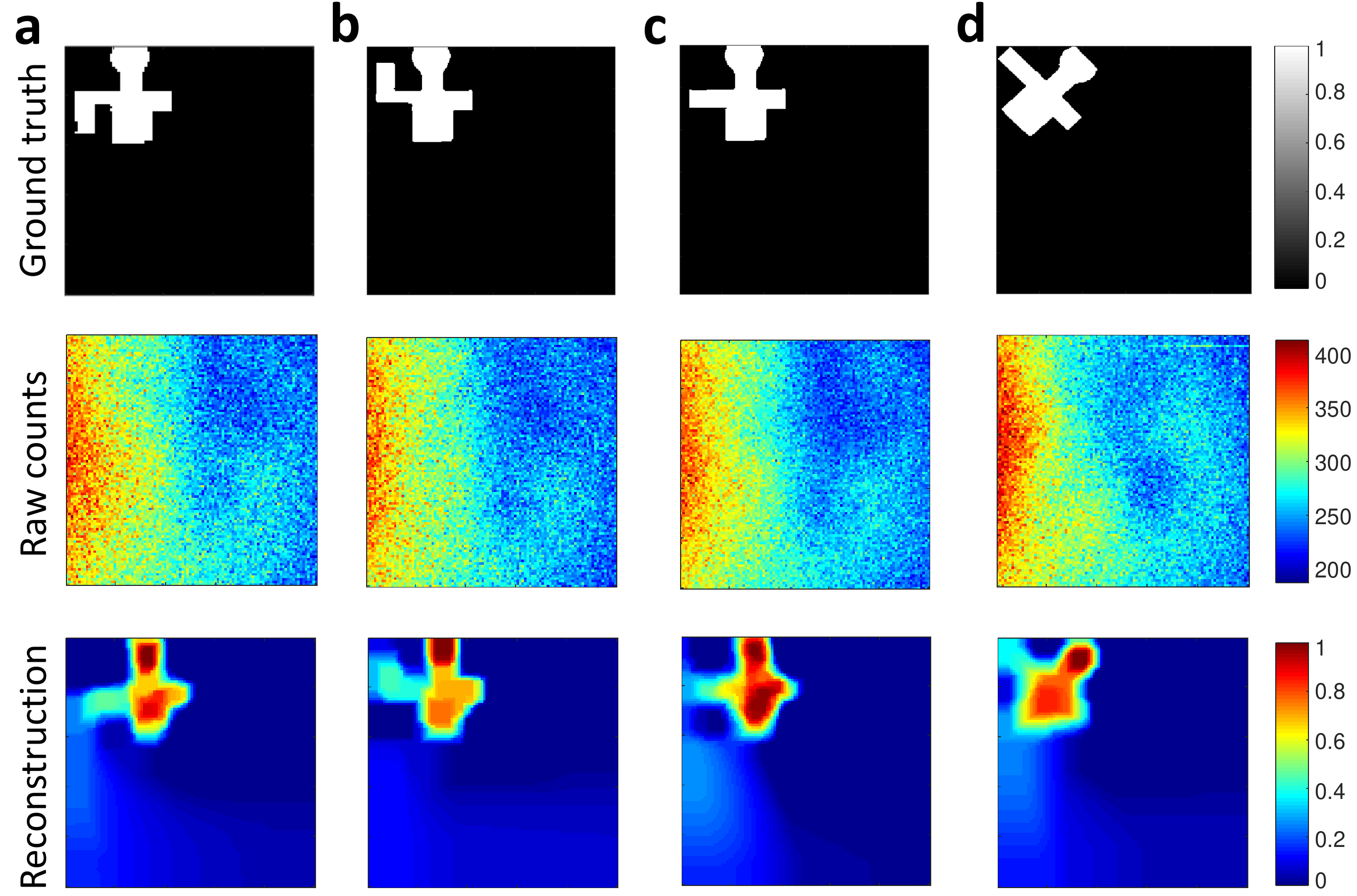}}
\caption{Experimental results on the recovery of different hidden-wall reflectivity patterns, (a)--(d). First row: ground  truth patterns on the hidden wall; second row: raw photon counts for   $100\times 100$ raster-scanned laser positions; third row:   reconstructions in the presence of the occluder, based on solving   Eq.~(\ref{eq:PML_general}), showing that detailed scene features are successfully recovered.  }
\label{Fig:patterns}
\end{figure}

First, we validate our occluder-assisted NLoS imaging method by
reconstructing different reflectivity patterns on the hidden
wall. These results are summarized in Fig.~\ref{Fig:patterns}. Four
reflectivity patterns were placed on the hidden wall, as shown in Fig.~\ref{Fig:patterns}'s first row. The laser's dwell time at each
raster-scanned point was set so that $N=7.12\times 10^{5}$ pulses
were sent, resulting in $\sim$275 detected photons per pixel (PPP) on
average. For each reflectivity pattern, a matrix of $100\times100$
raw counts was collected, as given in Fig.~\ref{Fig:patterns}'s middle row. The
reflectivity patterns on the hidden wall were then reconstructed using our
algorithm for solving Eq.~(\ref{eq:PML_general}), successfully
revealing their fine details, as seen in Fig.~\ref{Fig:patterns}'s bottom row.

\begin{figure}[tbh]
  \centering
  \centerline{\includegraphics[width=0.98\textwidth]{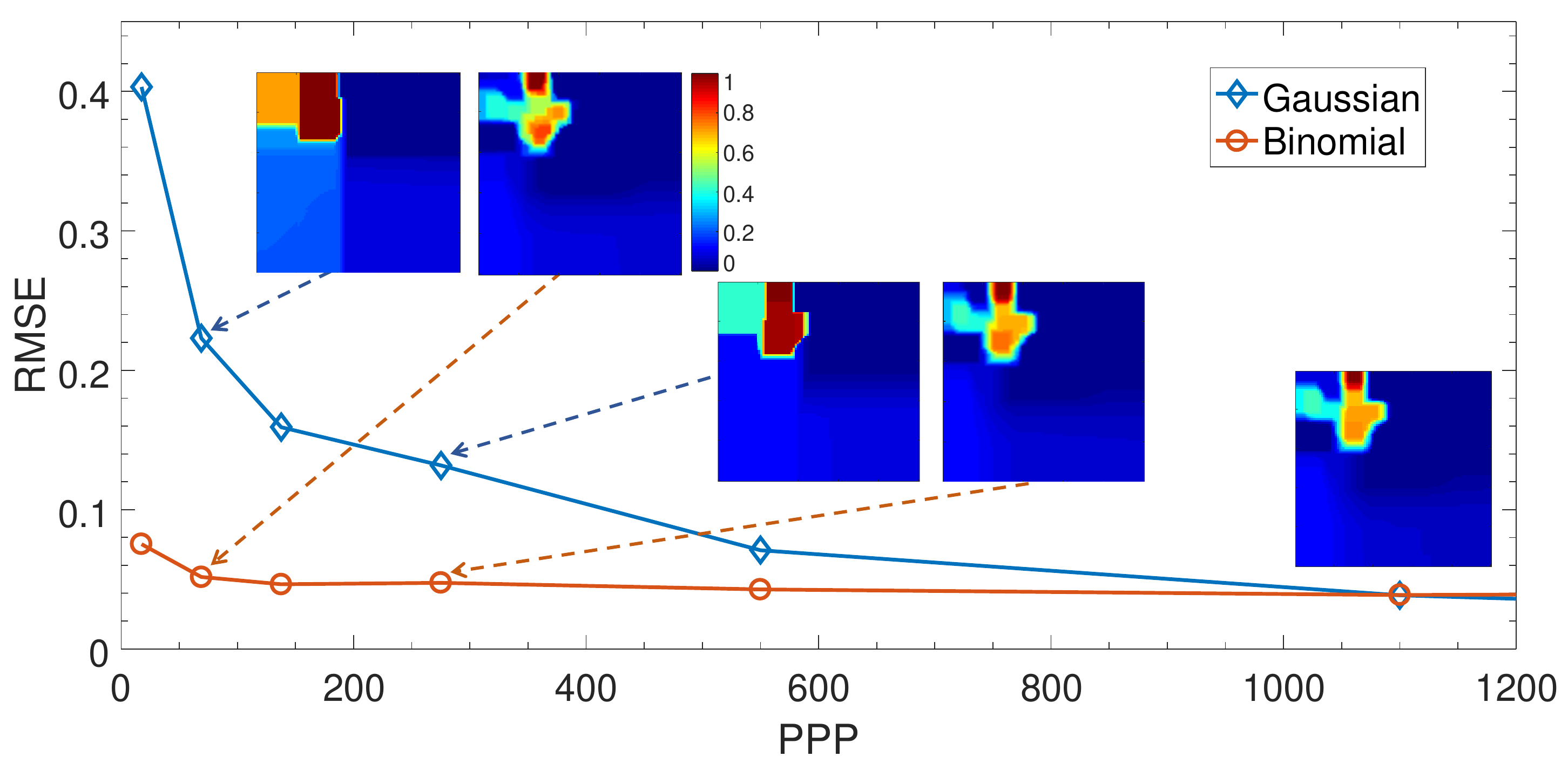}}
\caption{Root-mean-square error (RMSE) and reconstruction results (insets) with different numbers of detected photons per pixel (PPP). The RMSE of our binomial-likelihood method remains below 0.05 with $>$69 detected PPP, whereas the Gaussian-likelihood method employed in \cite{theory2017} requires at least $\sim$1100 detected PPP to achieve similar performance.  }
\label{Fig:PPP}
\end{figure}

\begin{figure}[tbh]
  \centering
  \centerline{\includegraphics[width=0.9\textwidth]{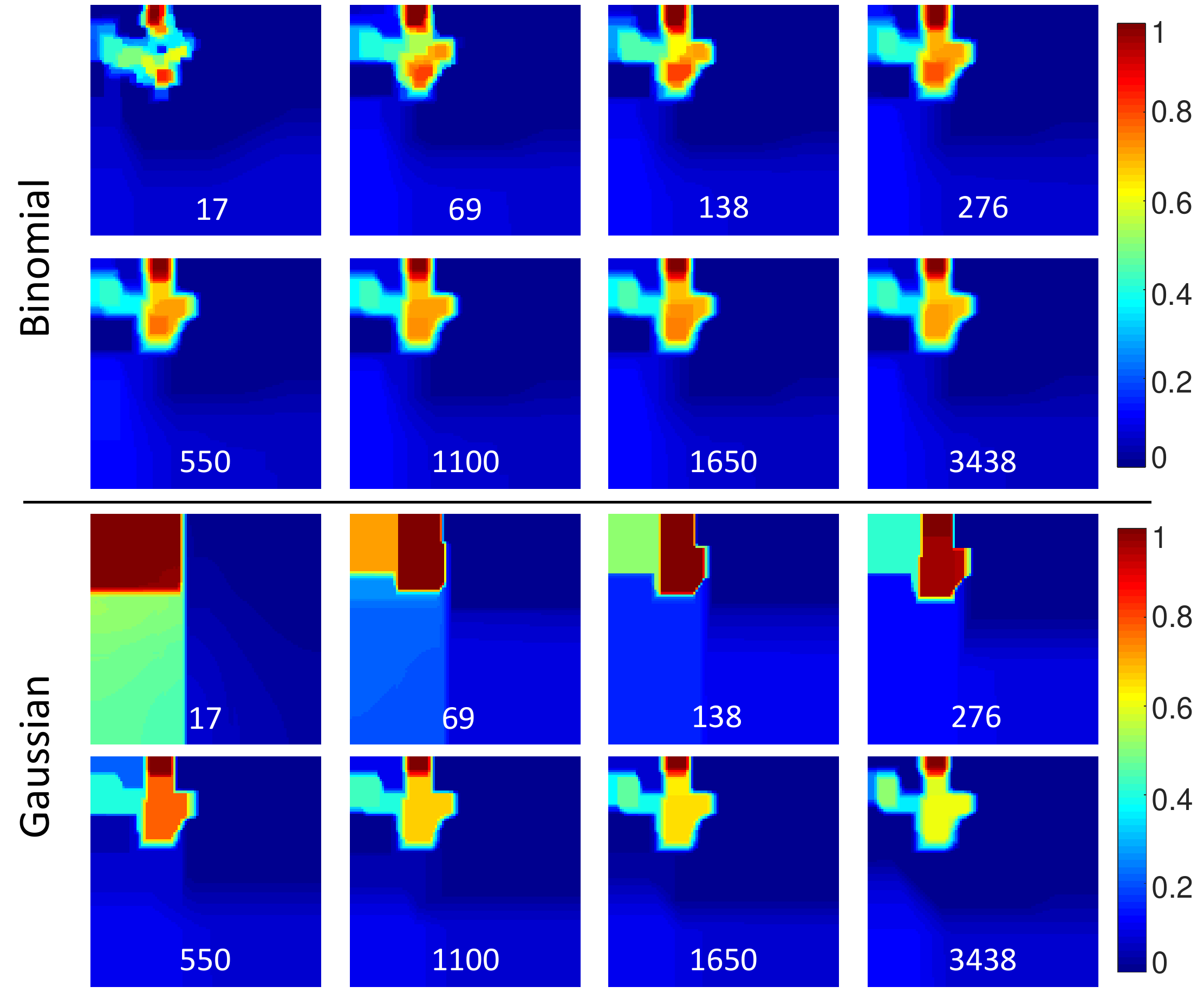}}
\caption{Reflectivity reconstructions with different numbers of detected photons per pixel (PPP). We compare the binomial-likelihood algorithm (Eq.~(\ref{eq:PML_general})) and the Gaussian-likelihood algorithm~\cite{theory2017} for different numbers of average detected PPP, ranging from 17 to 3438 as indicated on the bottom of each figure. The photon efficiency of the binomial-likelihood method is far superior to that of the Gaussian-likelihood method, with the latter requiring at least $\sim$1100 PPP to achieve reconstructions comparable to those of the former with $\sim$69 PPP. In the low-photon detection regime, PPP\,$<275$, the Gaussian-likelihood method fails to reconstruct the details of the reflectivity image.
} \label{fig:suppfig4}
\end{figure}

To quantify the photon efficiency and fidelity of our method, we varied
the dwell time per laser illumination point (which determines the overall
acquisition time) and tracked reconstruction performance as a function
of the empirical average PPP, as shown in Figs.~\ref{Fig:PPP} and~\ref{fig:suppfig4}. We measure the reconstruction fidelity by the root-mean-square error (RMSE) in the reconstructed reflectivity $\hat{{\bf F}}$,
\begin{equation}
{\rm RMSE}(\hat{{\bf F}},{\bf F}) = \frac{1}{n}
\vphantom{\sum_{k,l}} \sqrt{{\sum_{k,l}} \bigl( F_{kl} - \hat{F}_{kl} \bigr)^2}\,, \nonumber
\end{equation}
where ${\bf F}$ is the true reflectivity pattern as determined from measurements in the high
photon-count limit. It is evident from Fig.~\ref{Fig:PPP} that reconstruction fidelity for our
binomial-distribution-based likelihood method does not degrade much
(remains below 0.05) as the average PPP decreases from $\sim$1100 to $\sim$100. Figure~\ref{Fig:PPP} also shows RMSE for
the Gaussian-distribution-based likelihood method employed in \cite{theory2017}. We see that the binomial-likelihood method's photon
efficiency is substantially better than that of the Gaussian-likelihood method: the latter
requires at least $\sim$1100 detected PPP to achieve a fidelity
similar to what the former realized with only $\sim$69 detected PPP. This behavior is mainly due
to the mismatched noise model in the standard Gaussian-likelihood method,
which presumes the noise to be additive, signal independent, and
Gaussian distributed, whereas in the low photon-count regime without photon-number resolution it is really signal dependent and binomial distributed.

\begin{figure}[tbh]
  \centering
  \centerline{\includegraphics[width=0.95\textwidth]{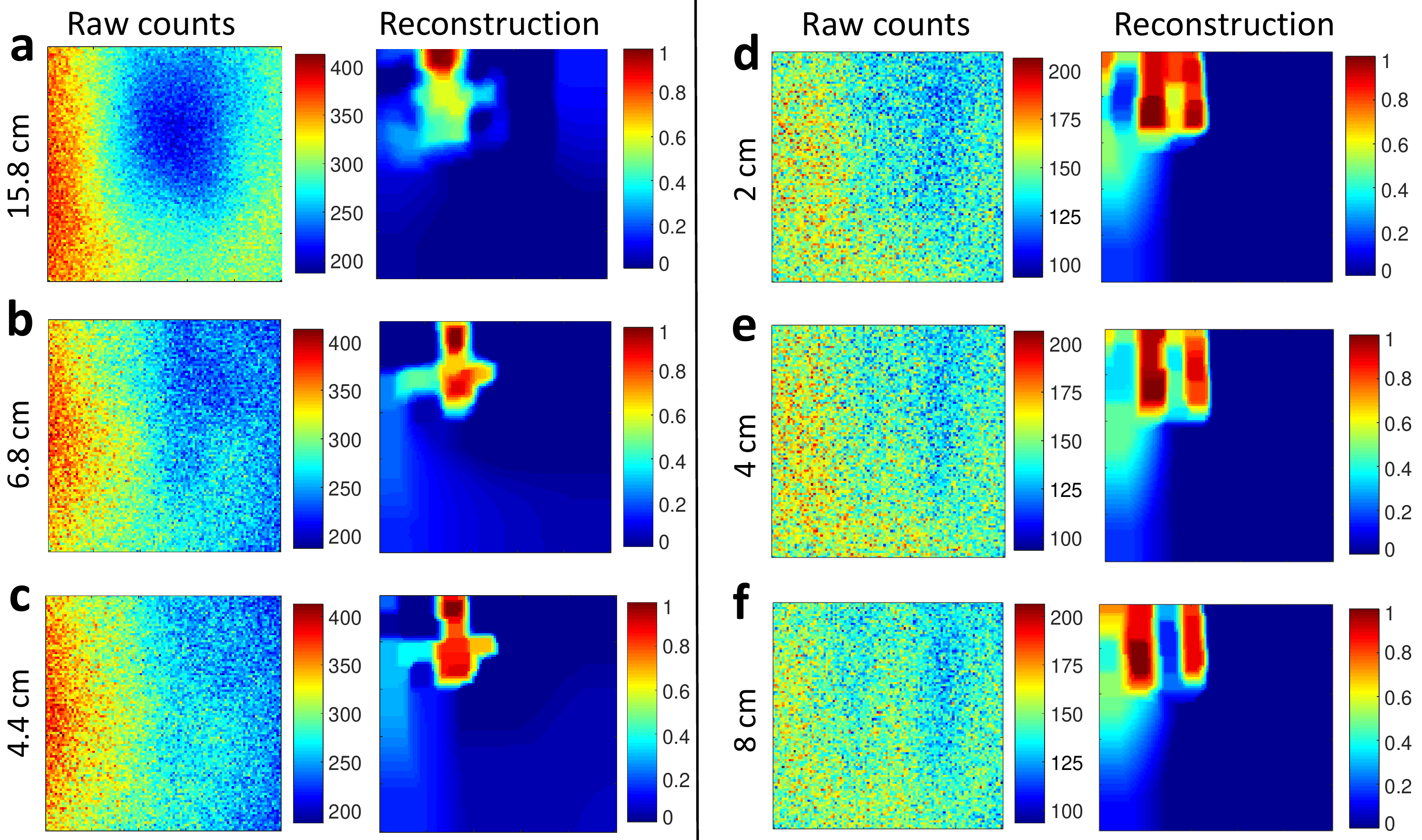}}
\caption{(a)--(c), Reconstructions
 of the Fig.~\ref{Fig:patterns}(a) reflectivity pattern obtained using circular occluders with
  diameters of 15.8\,cm, 6.8\,cm and 4.4\,cm. A small (large) occluder
  sharpens (blurs) the image. (d)--(f), Reconstructions of two-bar reflectivity patterns
 with bar separations of 2\,cm, 4\,cm and 8\,cm that were obtained using a 6.8-cm-diameter circular occluder. Our system achieves 4\,cm spatial resolution.}
\label{Fig:occluder}
\end{figure}

Finally, we quantify the effect of occluder size (Figs.~\ref{Fig:occluder}(a)--(c)) and the limits of achievable
spatial resolution (Figs.~\ref{Fig:occluder}(d)--(f)). In Figs.~\ref{Fig:occluder}(a)--(c), we used our system to image the
reflectivity pattern of Fig.~\ref{Fig:patterns}(a) using circular occluders, whose diameters ranged from 15.8\,cm to 4.4\,cm,
while keeping other experimental parameters unchanged. The results show that a small (large) occluder sharpens (blurs) the image, similar to conventional pinhole imaging. In Figs.~\ref{Fig:occluder}(d)--(f), we fixed the diameter of the circular occluder at 6.8\,cm and reconstructed a hidden-wall reflectivity pattern consisting of two bars with varying separation. With this occluder we see that our
system provides $\sim$4\,cm spatial resolution.

\section{Discussion}
\begin{figure}[tbh]
  \centering
  \centerline{\includegraphics[width=0.9\textwidth]{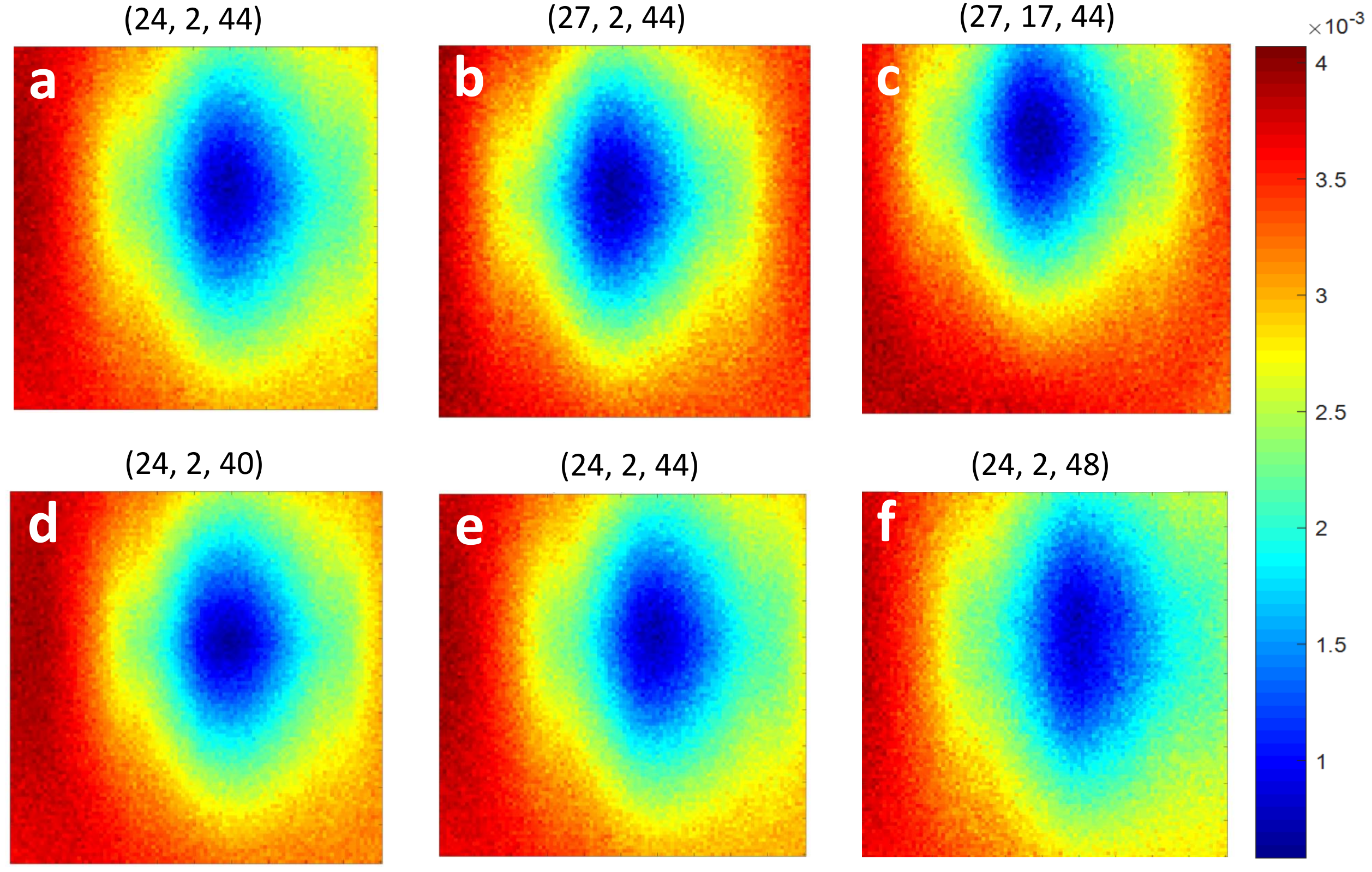}}
\caption{Raw detected-count measurements with a 15.8-cm-diameter occluder placed at different positions. The real location $(X,Y,Z)$ (cm) of the occluder is indicated on the top of each figure. In (a)--(c), we fixed the position of the occluder on the $Z$ axis and shifted it along the $X$ and $Y$ axes: the center of the rings reveals the $(X,Y)$ position of the occluder. In (d)--(f), we fixed the position of the occluder on the $X$ and $Y$ axes and shifted it along the $Z$ axis: the size of the rings reveals the $Z$-axis position of the occluder. These preliminary measurements suggest that occluder position may be localized from raw-count data.
} \label{fig:suppfig5}
\end{figure}

We have assumed throughout that the occluder's location was known and that it was nonreflecting. These assumptions may be relaxed.  In particular, the location of a nonreflecting occluder may be obtained using a blind deconvolution method~\cite{theory2017}, in which the occluder and the scene hidden behind it are reconstructed jointly. The viability of this approach is suggested by the fact that the occluder can be localized from the raw counts, as shown in Fig.~\ref{fig:suppfig5}. Furthermore, if the occluder has nonzero reflectivity, its contribution to the raw photon counts can be modeled using the principles employed in our forward model and incorporated into the blind deconvoloution procedure.

%

In conclusion, we have demonstrated a framework for photon-efficient, occluder-facilitated NLoS imaging. Our results may ultimately lead to new imaging methodologies capable of opportunistically exploiting diverse features of the environment---including, but not limited to, simple occluders---and thus pave the way to NLoS imaging in a wide variety of applications.

\section*{Funding}
Defense Advanced Research Projects Agency (DARPA) REVEAL Program Contract No.~HR0011-16-C-0030.

\section*{Acknowledgments}
The authors thank Changchen Chen for assistance with figure preparation, Connor Henley for the measurements reported in Fig.~11, and William T. Freeman and Vivek K Goyal for helpful discussions.

\section*{Appendices}

\appendix
\section{Theoretical details}~\label{App:theory}
\vspace*{-0.32in}

\paragraph{Light Propagation Model}
Here we provide details for the forward model in Eq.~(\ref{eq:linear_measurements_4d}). The ray-optics propagation model we use for third-bounce light is that from~\cite{theory2017}, which we present in detail for ease of reference.  Unlike~\cite{theory2017}, which assumes additive, signal-independent, Gaussian noise, our forward model accurately captures the noise statistics for SPAD detection in the low-photon-count regime.

Light propagates from the laser, located at position $\boldsymbol{\Lambda}$, until it reaches the detector, located at position $\boldsymbol{\Omega}$, while accounting for a three-bounce propagation path. Our goal is to reconstruct the reflectivity function $f(\x)$, for $\x \in \mathcal{S}$, where $\mathcal{S}$ is a two-dimensional parameterization of the hidden wall.

Figure~\ref{fig:suppfig1} illustrates a three-bounce trajectory of the form $\boldsymbol{\Lambda}\rightarrow\ellb_{ij}\rightarrow\x\rightarrow\cb\rightarrow\boldsymbol{\Omega}$, where $\ellb_{ij}$ is the $ij$th position in the laser's illumination grid, $\x$ is a point on the hidden wall, and $\cb$ is a point on the visible wall that is in the SPAD's field of view. For single-pulse illumination of $\ellb_{ij}$, the average number of photons following this trajectory that arrive at the SPAD is
\begin{equation}
K_{{\rm p}} f(\x)\,\frac{G_{\boldsymbol{\Lambda},\ellb_{ij},\x,\cb,\boldsymbol{\Omega}}\,{\rm d}\x\,{\rm d}\cb\,{\rm d}\boldsymbol{\Omega}}{\|\ellb_{ij}-\x\|^2\|\x-\cb\|^2\|\cb- \boldsymbol{\Omega} \|^2}\,, \nonumber
\end{equation}
where $K_{{\rm p}}$ is the average number of photons per pulse emitted by the laser, and ${\rm d}\x,{\rm d}\cb,{\rm d}\boldsymbol{\Omega}$ are differential areas. This expression accounts for the inverse-square-law losses experienced in free-space light propagation from $\ellb_{ij}$ to $\x$, from $\x$ to $\cb$, and from $\cb$ to $\boldsymbol{\Omega}$, as well as the linear scaling by $f(\x)$ that results from reflection at $\x$. The geometric factor $G_{\boldsymbol{\Lambda},\ellb_{ij},\x,\cb,\boldsymbol{\Omega}}$ combines the Lambertian bidirectional reflectance distribution functions (BRDFs) associated with the diffuse reflections at the visible wall and the hidden wall, and is given by
\begin{eqnarray}
G_{\boldsymbol{\Lambda},\ellb_{ij},\x,\cb,\boldsymbol{\Omega}}
\hspace*{-.1in}&\equiv&\hspace*{-.1in}\cos(\boldsymbol{\Lambda}-\ellb_{ij},{\bf n}_{\ellb_{ij}})\cos({\x}-\ellb_{ij},{\bf n}_{\ellb_{ij}}) \nonumber \\
&&\!\!\times\cos(\x-\ellb_{ij},{\bf n}_{\x})\cos(\x-\cb,{\bf n}_{\x})\cos(\x-\cb,{\bf n}_{\cb})\cos(\cb-\boldsymbol{\Omega},{\bf n}_{\cb})\,,\nonumber
\end{eqnarray}
where ${\bf n}_{\ellb_{ij}},{\bf n}_{\x},{\bf n}_{\cb}$ are the surface normals at $\ellb_{ij},\x, \cb$, respectively, and $\cos({\boldsymbol a},{\boldsymbol b})$ is the cosine of the angle between the vectors ${\boldsymbol a}$ and ${\boldsymbol b}$.

For single-pulse illumination of $\ellb_{ij}$, we use $Y_{ij}$ to denote the average number of photons arriving at the detector from three-bounce trajectories. Deriving an expression for $Y_{ij}$ entails summation over all such paths.  In particular, this means summing over: (1) all $\x\in\mathcal{S}(\ellb_{ij},\cb)$, where $\mathcal{S}(\ellb_{ij},\cb)$ is the section of the hidden wall $\mathcal{S}$ that has an unoccluded line of sight to both $\ellb_{ij}$ and $\cb$; (2) all $\cb\in\mathcal{C}$, where $\mathcal{C}$ is a parameterization of the section of the visible wall that is in the SPAD's field of view; and (3) all points in $\mathcal{D}$, the SPAD detector's photosensitive region. With these definitions we then have:
\begin{equation} \label{eq:meas_fwd_model}
Y_{ij}= K_{{\rm p}} \int\limits_{\mathcal{S}(\ellb_{ij},\cb)}\!\mathrm{d}\x \int\limits_{\mathcal{C}}\!\mathrm{d}\cb\int\limits_{\mathcal{D}}{\rm d}\boldsymbol{\Omega}\,f(\x) \frac{G_{\boldsymbol{\Lambda},\ellb_{ij},\x,\cb,\boldsymbol{\Omega}}}{\|\ellb_{ij}-\x\|^2\|\x-\cb\|^2\|\cb- \boldsymbol{\Omega} \|^2}\,,
\end{equation}
where Eq.~(\ref{eq:meas_fwd_model})'s spatial integrations account for all possible three-bounce trajectories from the laser to the detector. This result can be simplified as follows:
\begin{eqnarray} \label{eq:meas_fwd_model_simple}
Y_{ij}\hspace*{-.1in}&=&\hspace*{-.1in} K_{{\rm p}} \int\limits_{\mathcal{S}}\!\mathrm{d}\x\,f(\x)\int\limits_{\mathcal{C}}\!\mathrm{d}\cb\int\limits_{\mathcal{D}}{\rm d}\boldsymbol{\Omega}\, \frac{{1}_{\mathcal{S}(\ellb_{ij},\cb)}(\x)G_{\boldsymbol{\Lambda},\ellb_{ij},\x,\cb,\boldsymbol{\Omega}}}{\|\ellb_{ij}-\x\|^2\|\x-\cb\|^2\|\cb- \boldsymbol{\Omega} \|^2}\nonumber\\
\hspace*{-.1in}&=&\hspace*{-.1in}K_{{\rm p}}\int\limits_{\mathcal{S}}\! \mathrm{d}\x\,f(\x)A^{(ij)}(\x)\,,
\end{eqnarray}
where $1_{\{\x' \} }(\x)$ is the indicator function (i.e., it equals 1 if and only if $\x=\in \{\x'\}$ and is 0 otherwise), and for fixed $i,j$ we have defined
\begin{equation}\label{eq:A_def}
A^{(ij)}(\x)\equiv \int\limits_{\mathcal{C}}\!\mathrm{d}\cb\int\limits_{\mathcal{D}}{\rm d}\boldsymbol{\Omega}\,\frac{{1}_{\mathcal{S}(\ellb_{ij},\cb)}(\x)G_{\boldsymbol{\Lambda},\ellb_{ij},\x,\cb,\boldsymbol{\Omega}}}{\|\ellb_{ij}-\x\|^2\|\x-\cb\|^2\|\cb- \boldsymbol{\Omega} \|^2}\,.
\end{equation}
Equation~(\ref{eq:meas_fwd_model_simple}) is specified in terms of the continuous variable $\x$. In what follows, it will be convenient to discretize the coordinate system on the hidden wall $\mathcal{S}$ by introducing an $n\times n$ grid indexed by $(k,l)$. We then have that $A^{(ij)}(\x)$ becomes $A^{(ij)}_{kl}$ and $f(\x)$ becomes $F_{kl}$. Making these substitutions in (\ref{eq:meas_fwd_model_simple}) we obtain the discrete version of the forward model that appeared in Eq.~(\ref{eq:linear_measurements_4d}):
\begin{equation} \label{eq:linear_measurements}
Y_{ij}= K_{{\rm p}}\sum\limits_{k,l}A^{(ij)}_{kl}F_{kl}\,.
\end{equation}

\paragraph{Shadow Function}
Equations~(\ref{eq:A_def}) and (\ref{eq:linear_measurements}) show that the presence of an occluder only affects $Y_{ij}$ through its impact on $\mathcal{S}(\ellb_{ij},\cb)$, i.e., the patch on the hidden wall that has unobstructed lines of sight to both $\ellb_{ij}$ and $\cb$. To better understand this connection between the occluder and $\mathcal{S}(\ellb_{ij},\cb)$, we introduce a binary \emph{shadow function} $\Theta(\x,\y)$ that indicates whether point $\x$ on the hidden wall and point $\y$ on the visible wall are \emph{visible} to each other:
\begin{equation}\label{eq:vis}
\Theta(\x,\y)=\left\lbrace
\begin{array}{ll}
1, & \mbox{unobstructed line of sight between $\x$ and $\y$,}\\
0, & \mbox{obstructed line of sight between $\x$ and $\y$.}
\end{array}
\right.
\end{equation}
With this definition we have $\mathcal{S}(\ellb_{ij},\cb) = \left\{\x \in \mathcal{S} : \mbox{$\Theta$}(\x,\ellb_{ij})\mbox{$\Theta$}(\x,\cb)=\mbox{1} \right\}$, i.e., it is the subset of hidden-wall positions $\mathcal{S}$ that satisfy both ${\Theta(\x,\ellb_{ij})}=1$ and ${\Theta(\x,\cb)}=1$. Note that $\mathcal{S}(\ellb_{ij},\cb)$ and $\mathcal{S}(\ellb_{i'j'},\cb)$ differ on hidden-wall patches for which the occluder blocks light from $\ellb_{ij}$ but not from $\ellb_{i'j'}$ or vice versa.

\paragraph{Informative Measurements}
Our experiment raster scans the grid points $\{\ellb_{ij}\}$ on the visible wall and detects third-bounce photons reflected from a large portion of that wall.  The informativeness of these measurements stems from the diversity of the coefficients $A^{(ij)}_{kl}$.  In the absence of an occluder, we have $\mathcal{S}(\ellb_{ij},\cb)=\mathcal{S}$ for all $\ellb_{ij}$ and $\cb$. From (\ref{eq:A_def}) we then see that the dependence of $A_{kl}^{(ij)}$ on $i,j$ originates from the product of two smoothly-varying functions---the inverse-square-law term $\|\ellb_{ij}-\x_{kl}\|^{{-}2}$ and the geometric function $G_{\boldsymbol{\Lambda},\ellb_{ij},\x_{kl},\cb,\boldsymbol{\Omega}}$---that yield smooth variations in $A_{kl}^{(ij)}$ as $(i,j)$ changes.  In the presence of an occluder, however, the impact of nontrivial shadow functions in determining $\mathcal{S}(\ellb_{ij},\cb)$ makes $A_{kl}^{(ij)}$ vary more abruptly with $(i,j)$ changes, greatly increasing the informativeness of the measurements.

\begin{figure}[tbh]
	\centering
	\centerline{\includegraphics[width=0.7\textwidth]{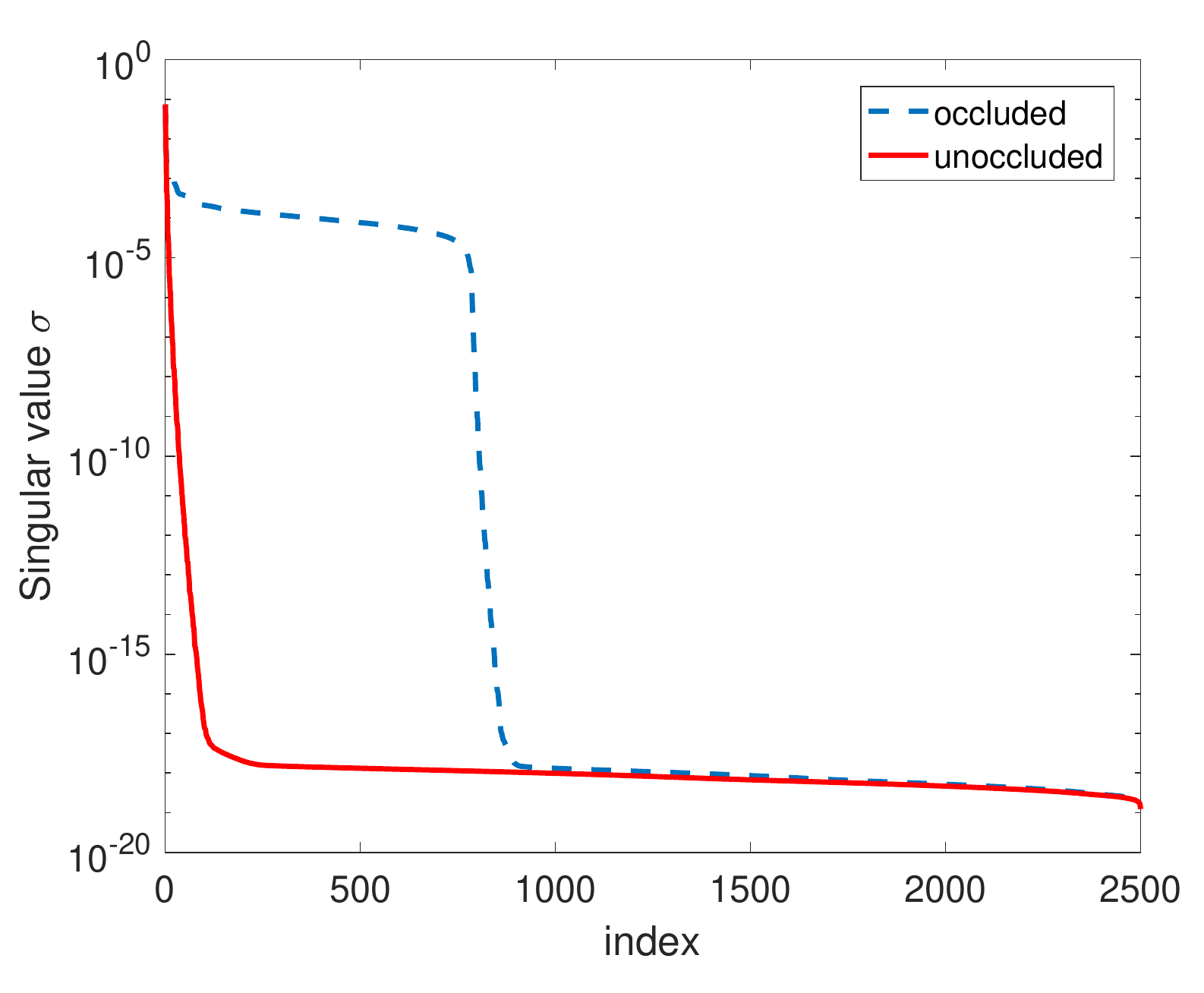}}
	\caption{Comparing the informativeness of occluded and unoccluded measurements. We numerically simulated the setup of Fig.~\ref{fig:suppfig1} and evaluated the informativeness of the measurements with and without an occluder from the $\A$ matrix's singular values $\{\sigma\}$. In our simulations, the laser illuminates a $50\times 50$ grid on the visible wall, and the hidden wall is discretized to a $50\times 50$ grid. The singular values of the corresponding $2500\times 2500$ ${\bf A}$ matrix were calculated for an occluded setup (blue dashed curve) and an unoccluded setup (red solid curve). The singular values of the occluded ${\bf A}$ matrix are substantially higher than those of the unoccluded matrix, suggesting that measurements in the occluded setup will be much more informative.} \label{fig:suppfig_svd}
\end{figure}

To demonstrate this effect, we rearrange $\left\lbrace Y_{ij} \right\rbrace$ as an $m^2$-dimensional column vector ${\bf y}$, $\left\lbrace F_{kl} \right\rbrace$ as an $n^2$-dimensional column vector ${\bf f}$, and $\{\A^{(ij)}\}$ as an $m^2\times n^2$-dimensional matrix ${\bf A}$ such that Eq.~(\ref{eq:linear_measurements_4d}) for $1\le i,j\le m$ gets combined into
\begin{equation}
{\bf y}={\bf A}{\bf f}.
\end{equation}
With this rearrangement we can evaluate the informativeness of the measurements by analyzing the spectral properties of ${\bf A}$. Toward that end, Fig.~\ref{fig:suppfig_svd} shows the singular values of ${\bf A}$ for two experimental setups. The first setup corresponds to an unoccluded scene, whereas the second setup corresponds to an occluded scene, in which a black circular patch has been inserted between the visible and hidden walls. It is evident from these singular values that the occluded measurements are substantially more informative, suggesting that the presence of the occluder will enable higher-fidelity reconstruction of the hidden-wall's reflectivity pattern.

\paragraph{Measurement Statistics}
The laser illuminates position $\ellb_{ij}$ with $N$ pulses before it addresses the next grid point on the visible wall. Each pulse that illuminates $\ellb_{ij}$ results in an average of $Y_{ij}$ third-bounce photons arriving at the detector's location $\boldsymbol{\Omega}$.  In this \emph{low-flux} regime the number of detections registered by a photon-number resolving detector from illumination of $\ellb_{ij}$ by a single pulse is Poisson distributed, with mean $\eta(Y_{ij}+B_{ij})$, where $\eta$ is the detector's quantum efficiency, and $B_{ij}$ is the average number of background-light photons arriving during a single-pulse measurement interval (see details below). A SPAD detector, however, is not number resolving; it suffers a dead time after making a single detection that, for our experiment, precludes more than one detection in a single-pulse measurement interval. In this case, each optical pulse can yield either a 0 count or 1 count, and these events occur with probabilities $P_0^{(ij)}({\bf F})$ and $1-P_0^{(ij)}({\bf F})$, respectively, where
\begin{equation}\label{eq:Poisson_meas}
P_0^{(ij)}({\bf F}) = \exp[-\eta( Y_{ij}+B_{ij})] \approx 1-\eta(Y_{ij}+B_{ij})\,,
\end{equation}
The equality in Eq.~(\ref{eq:Poisson_meas}) comes from the Poisson distribution. The approximation for that Poisson probability is due to the enormous attenuation incurred in three diffuse reflections, which makes $\eta Y_{ij} \ll 1$, and the pre-detection optical filtering used to ensure that $\eta B_{ij} \ll 1$, which prevents SPAD counts from occurring in every single-pulse measurement interval.  The ${\bf F}$ dependence of $P_0^{(ij)}({\bf F})$ arises from the $Y_{ij}$ term, see Eq.~(1).

The statistical independence of the photon counts from different laser pulses now makes $R_{ij}$, the total photon count from the $N$ pulses that illuminate $\ellb_{ij}$, a binomial random variable with success probability $1-P_0^{(ij)}({\bf F})$, i.e.,~\cite{shin2015TCI}
\begin{equation}\Pr(R_{ij};{\bf F}) = \left(\begin{array}{c}N \\ R_{ij}\end{array}\right)[1-P_0^{(ij)}({\bf F})]^{R_{ij}}[P_0^{(ij)}({\bf F})]^{N-R_{ij}}.
\end{equation}
Using this binomial distribution for the count statistics, and dropping terms that are independent of ${\bf F}$, we get the following negative log-likelihood function for the raw count matrix ${\bf R}$ given the reflectivity matrix ${\bf F}$:
\begin{eqnarray} \label{eq:minus_log_likelihood}
&-\log[\mathcal{L}({\bf R};{\bf F} )]=-\log\!\left(\prod\limits_{i,j}\Pr(R_{ij}; {\bf F})\right)=-\sum\limits_i\sum\limits_j \log\!\left[\Pr(R_{ij};{\bf F})\right] \\
&=\sum\limits_i\sum\limits_j \left\lbrace (N-R_{ij}) \left[ \eta  K_{{\rm p}}  \sum\limits_{k,l}A^{(ij)}_{kl}F_{kl} \right]  -R_{ij} \log \left[  \eta K_{{\rm p}} \sum\limits_{k,l}A^{(ij)}_{kl}F_{kl} + \eta B_{ij} \right]  \right\rbrace,
\end{eqnarray}
where the first equality in (\ref{eq:minus_log_likelihood}) follows from the statistical independence of the shot noises generated by different laser pulses.

\section{Experimental details}~\label{App:experiment}
\vspace*{-.32in}

\begin{figure}[tbh]
  \centering
  \centerline{\includegraphics[width=0.9\textwidth]{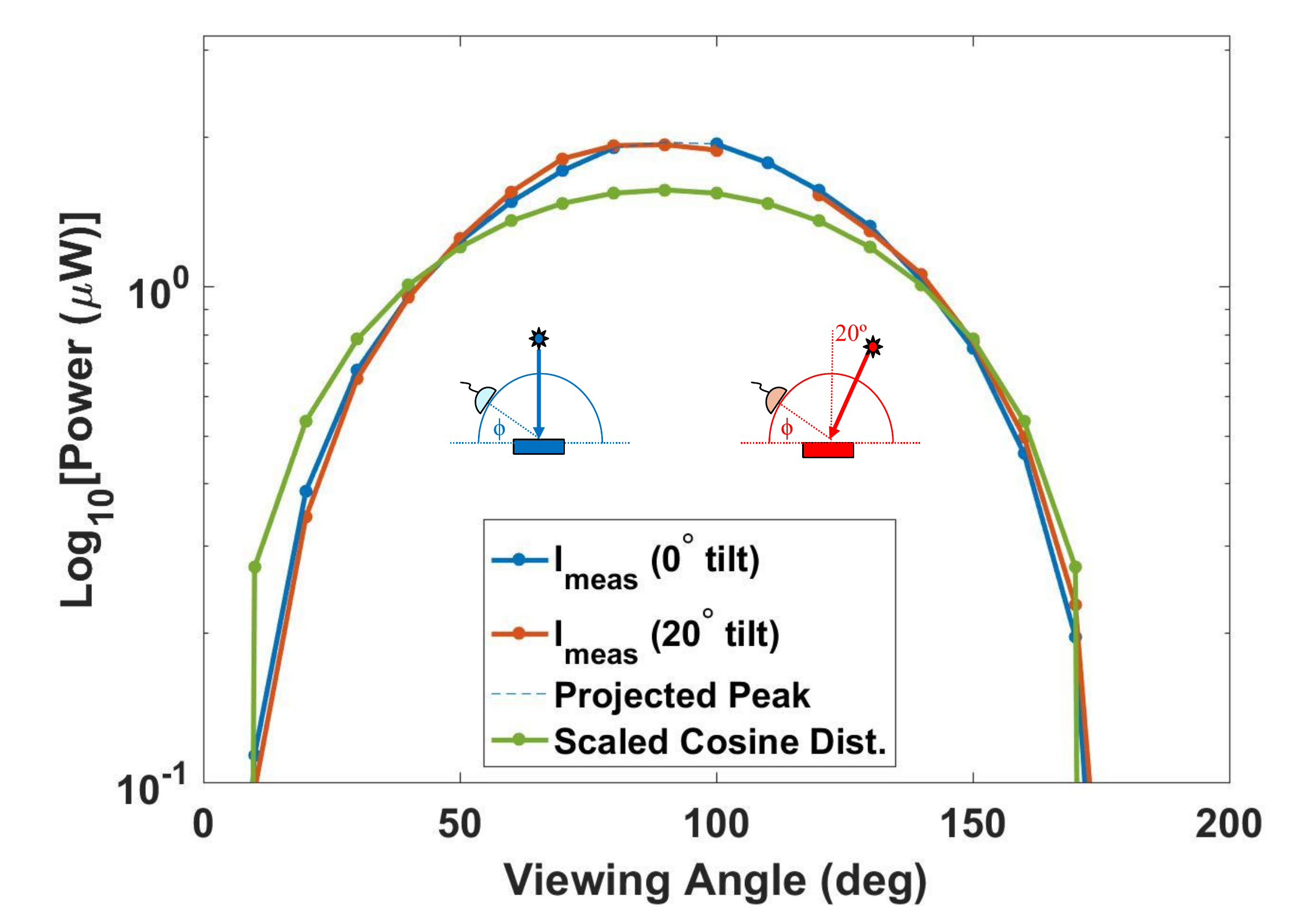}}
\caption{Near-Lambertian reflectance behavior of white poster-board visible wall. The blue (red) data points correspond to measurements made with the setup in the blue (red) inset: a laser illuminated the visible wall at normal incidence (20$^{\circ}$ offset from normal incidence), and a detector recorded the power reflected at different viewing angles. The green line is the theoretical cosine curve for a perfect Lambertian surface. We find that the visible wall has $\sim$80\% reflectivity and is nearly Lambertian except for a small specular component when the viewing angle is perpendicular to the surface. We also performed this characterization for the patterns on the hidden wall, and found that the Lambertian property of those patterns was similar to that of the visible wall.
} \label{fig:suppfig6}
\end{figure}

\paragraph{Visible wall characterization} We used a white poster board as a near-Lambertian reflecting surface to serve as the visible wall in our NLoS imaging experiment of Fig.~\ref{fig:suppfig1}. We used a 635-nm laser to illuminate the white board at two different incident angles and measured its reflected power at various viewing angles. The results are displayed in Fig.~\ref{fig:suppfig6}, showing that the white poster board is indeed nearly Lambertian.

\begin{figure}[tbh]
  \centering
  \centerline{\includegraphics[width=0.7\textwidth]{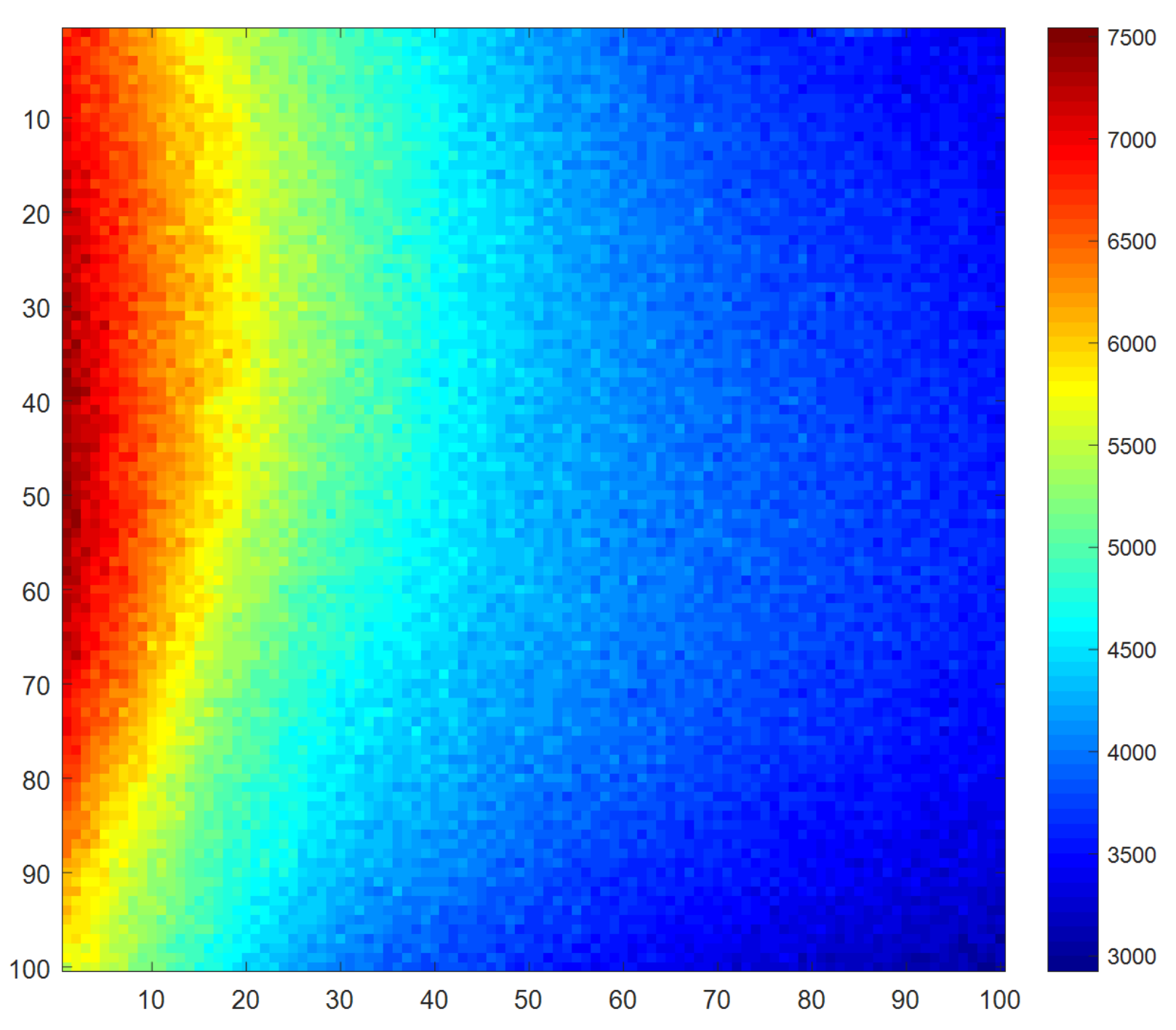}}
\caption{Results of long acquisition time background-light measurement used to calibrate $B_{ij}$. The reflectivity pattern on the hidden wall was replaced with a black surface. A total of 35.6 million laser pulses were transmitted at each laser point $\ellb_{ij}$ on the $100\times 100$ illumination grid, and the third-bounce counts were recorded by the SPAD. The nonuniformity is mainly due to scattering from the raster-scan galvo mirrors and SPAD afterpulsing that arises from detections of those first-bounce photons.} \label{fig:suppfig3}
\end{figure}

\paragraph{Background light} In Fig.~\ref{fig:suppfig3}, we show results of background-light detection over a long data-acquisition time that we used to calibrate $B_{ij}$, the average number of background photons arriving at the detector in a single-pulse measurement interval. For this measurement, the reflectivity pattern on the hidden wall was replaced with a black surface, a total of $N=3.56 \times 10^{7}$ laser pulses were transmitted at each laser point $\ellb_{ij}$, and the third-bounce photons were detected by the SPAD. Note that once performed this calibration applies to all subsequent measurements: in post-processing, we scale these background noise counts according to the dwell time used. The nonuniformity of the background counts is mainly due to scattering from the raster-scan galvo mirrors and to SPAD afterpulsing originating from detections of those first-bounce photons. Galvo-related background counts could be avoided with better scanning mirrors.

\end{document}